\documentclass[preprint,preprintnumbers,amsmath,amssymb]{revtex4}
\usepackage{graphicx}

\begin{document}


\title{ Radiation  in  Systems with Near Zero 
Dielectric Constant     }

\author{V.H.Arakelian $^{1}$ and  Zh.S.Gevorkian $^{1,2,\ast}$}

\affiliation{$^{1}$ Institute of Radiophysics and
Electronics,Ashtarak-2,0410, Armenia}

\affiliation{$^{2}$ Yerevan Physics Institute, Alikhanian Brothers
St. 2, Yerevan 0036, Armenia.\\
$^\ast$ Corresponding author:gevork@yerphi.am}



\date{\today}

\begin{abstract}
We discuss radiation effects in systems with near zero
dielectric constant. An enhancement of intensity by example 
of transition and diffusive radiations is demonstrated. A sharp
peak in the transition radiation spectral-angular 
intensity is predicted. Numerical estimates for  specific systems
are made and possible applications are discussed. 


\end{abstract}


\maketitle

{\it Introduction.}
Recently, systems with near zero dielectric permittivity(ENZ) 
become a subject of investigation because of their interesting
 properties \cite{Ziol,SeEn, ASSE}. One can find such materials at
 teraherz, infrared and optical frequencies among semiconductors,
 metals as well as dielectrics.
In the present paper we consider some radiation effects in the 
systems with near zero $\varepsilon$. The main underlying physical
idea is that in such media the pseudophoton momentum 
$k=\omega\sqrt{\varepsilon}/c$ is
small and therefore  is scattered on the inhomogeneties more 
effectively. Hence one can expect an enhancement of intensity in
radiation mechanisms caused by scattering of pseudophotons.
Below we discuss two such radiation mechanisms: transition(TR)
and diffusive(DR). In former case TR from a charged particle
traversing an interface between vacuum and an ENZ medium is 
investigated. In latter case we consider DR from a charged 
particle crossing a random stack with near zero average 
dielectric constant. 

{\it Transition Radiation.} 
Transition Radiation  originates when a charged particle 
crosses an interface between two homogeneous media \cite{GTs}. 
Its intensity depends on dielectric constants of media 
as well as on  particle velocity. In the simplest case 
of traversing of particle from vacuum into a ENZ medium the
backward spectral-angular TR intensity into vacuum is
determined as follows
\cite{GTs}
\begin{equation}
I(\omega,\theta)=\frac{2e^2\beta^2}{\pi c}\sin^3\theta 
\cos^2\theta\left|\frac{\varepsilon(\omega)-1}{\varepsilon(\omega)
\cos\theta+\sqrt{\varepsilon(\omega)-\sin^2\theta}}\right|^2
\label{TR}
\end{equation}
where $\beta=v/c$, $\theta$ is the angle between velocity directed
normally to the interface and observation direction, 
$\varepsilon(\omega)$ is the dielectric permittivity of a
homogeneous medium. Note that Eq.(\ref{TR}) is obtained from the
basic expression for TR intensity \cite{GTs} in the limit $\beta\ll1$
and $\varepsilon\ll 1$ . Relativistic case $\beta\sim 1$ will
 be investigated elsewhere. It follows from Eq.(\ref{TR}) that
the intensity has peak at angles
\begin{equation}
\sin^2\theta=\varepsilon(\omega) 
\label{angle}
\end{equation}
Reflection coefficient of a photon polarized in the plane of 
incidence from the interface between vacuum and $\varepsilon$-
medium is determined as \cite{LanLif}
\begin{equation}
r= \left(\frac{\varepsilon(\omega)\cos\theta-
\sqrt{\varepsilon(\omega)-\sin^2\theta}}
{\varepsilon(\omega)\cos\theta+\sqrt{\varepsilon(\omega)-
\sin^2\theta}}\right)^2
\label{reflec}
\end{equation}

As it follows from Eqs.(\ref{angle}) and (\ref{reflec}) 
maximum of TR intensity is achieved at the angles for which
the corresponding reflection coefficient is maximal $r=1$.  The
incident on the interface pseudophotons fulfilling the condition
Eq.(\ref{angle}) are completely reflected from the surface, 
separate from the charge and convert into real photons. Reflection 
coefficient Eq.(\ref{reflec}) becomes minimal $r=0$ 
at Brewster's angle\cite{LanLif}

\begin{equation}
tg^2\theta=\varepsilon(\omega)
\label{min}
\end{equation}
Pseudophotons fulfilling the condition Eq.(\ref{min}) does not
reflect and separate from the charge. Therefore corresponding TR
intensity is minimal for the angles and frequencies satisfying the
 Eq.(\ref{min}). These two values for ENZ medium are close to each
  other.
 Therefore one will has a sharp peak in spectral-angular TR intensity,
 see Fig.1. The width of peak on angle can be estimated as
 \begin{equation}
 \Delta\theta=arcsin\sqrt{\varepsilon}-arctg\sqrt{\varepsilon}
 \backsimeq \varepsilon^{3/2}
\label{width}
\end{equation}

In conventional media an ENZ conditions can be created in 
vicinity of frequencies of lattice longitudinal vibrations(optical
phonons) or collective oscillations of ensemble of free 
electrons(plasmons). The above mentioned peak is associated with
the properties of the medium and does not depend on the particle
energy. Its height for ENZ medium significantly exceeds
the maximum of conventional TR intensity in the limit $|\varepsilon|\to\infty  $ \cite{GTs}
\begin{equation}
I_0(\omega,\theta)\backsimeq\frac{4e^2\beta^2}{3\pi c}
\label{max}
\end{equation}

\begin{figure}
\includegraphics[width=8.4cm]{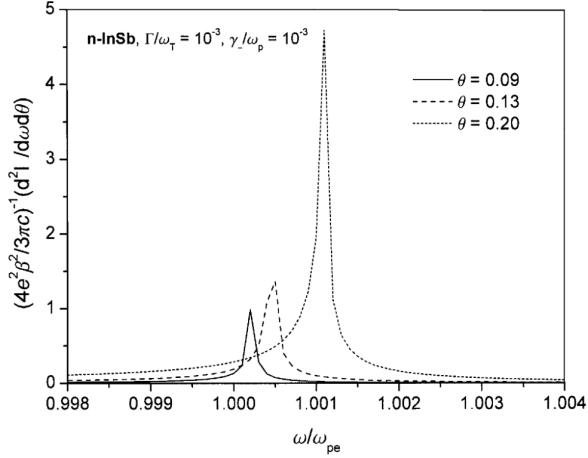}
\caption {Spectral-angular intensity of TR from a 
charged particle crossing the interface between vacuum and 
weakly doped semiconductor {\it InSb},$\varepsilon_0=17.72$,
$\varepsilon_{\infty}=15.68$, $\omega_p=1.9\times 10^{12}s^{-1}$.}
\label{fig.1}
\end{figure}

This enhancement, see Fig.(\ref{fig.1}) is caused by the resonance
 interaction of 
the electromagnetic field induced by a charged particle with
the vibrations of a lattice or with collective oscillations
of a plasma of free electrons. During this interaction a part
of energy of lattice vibrations or plasma oscillations is
transformed into the energy of transversal electromagnetic
wave. We make numerical estimates using Eq.(\ref{TR}) and the
following expression for the dielectric constant at frequencies 
close to the plasmon  frequencies of weakly doped {\it InSb}
 \cite{Agr}
\begin{equation}
\varepsilon(\omega)=\varepsilon_0\left(1-\frac{\omega_p^2}{
\omega^2(1+\frac{\gamma_-}{\omega})^2}\right)+i\left(\frac{
\varepsilon_0\gamma_-}{\omega^3(1+\frac{\gamma_-}{\omega})^2}+
(\varepsilon_0-\varepsilon_{\infty})\frac{\omega\Gamma}
{\omega_T^2}\right)
\label{eps}
\end{equation}
where $\varepsilon_{\infty}$ and $\varepsilon_0$ are high 
frequency and static dielectric constants,respectively,
$\omega_T$ is the transversal optical phonon frequency,$  \gamma_-$
and $\Gamma$ are the plasmon and  optical phonon decay rates,
respectively. As follows from 
Eq.(\ref{eps}), at frequencies close to plasmon 
frequency $\omega\sim\omega_p$, one can reach small values of
real and imaginary parts of dielectric 
constant  provided that $\gamma_-/\omega\ll 1$ and
$\Gamma/\omega_T\ll 1$.
As follows from Fig.(\ref{fig.1}) TR photons from the 
interface between vacuum and {\it InSb} belong to THz region.
In Fig.1 TR intensity from ENZ medium is normalized to the
 TR intensity value from a medium with $|\varepsilon|\to\infty$.
 Therefore TR intensity from ENZ medium exceeds TR from a 
 metal almost $4-5$ times. Another advantage is that angular
 distribution is quite narrow compared to the metal case. 
 Hence TR from a ENZ medium can be considered as a promising 
 source of THz radiation.
 
{\it Diffusive Radiation.} A charged particle passing through
 a stack of plates placed in a homogeneous medium is known to
 be radiating electromagnetic waves. Radiation originates because
 of the scattering of electromagnetic field on the plates.
 Considering this problem theoretically it was shown 
 \cite{ZH98,ZH06} that when the conditions of diffusional 
 scattering of pseudophotons are fulfilled its contribution to
 the radiation intensity is the dominant one
\begin{equation}
I_D(\theta,\omega)=\frac{5e^2\gamma^2}{2\varepsilon
c}\frac{l_{in}^2(\omega)}{l^2(\omega)}\sin^2\theta
\exp\left[-\left(\frac{l}{l_{in}}\right)^{1/2}\frac{1}{|\cos\theta|}\right]
 \label{diff}
\end{equation}
where $\theta$ is the observation angle, 
$\vec v$ is the particle velocity which is directed normally to the
plates on $z$, $\varepsilon$ is the average dielectric constant
(see below), $\gamma=(1-v^2\varepsilon/c^2)^{-1/2}$ is the Lorentz factor
in the medium, $l,l_{in}$ are pseudophoton elastic and inelastic
mean free paths,respectively.  Inelastic mean
free path is mainly associated with the absorption of
electromagnetic field in the medium. Elastic mean free path is
associated with the photon refraction on plates. It depends on the
photon incidence angle on plates. In case of normally incident photon
elastic mean free path is determined as follows \cite{ZH98}
\begin{equation}
l=\frac{4k^2}{B(0)+B(2k)}
\label{elmean}
\end{equation}

where  $B$ is the correlation function of random dielectric
constant field created by randomly located plates 
\begin{equation}
B(q_z)=\frac{4(b-\varepsilon)^2n\sin^2q_za/2}{q_z^2}
\frac{\omega^4}{c^4}.
\label{corfun}
\end{equation}

Here $n=N/L_z$ is concentration of  plates in the system, $a$ is
their thickness, $b$ is their dielectric constant and $L_z$ is
is the system size in the $z$ direction.
 It follows from Eq.(\ref{corfun}) that when $ka\gg 1$,
$B(2k)/B(0)\sim 1/(ka)^2\ll 1$. Therefore  in both cases $ka\gg 1$
and $ka\ll 1$ photon mean free path has the form
\begin{equation}
l\sim \frac{k^2}{B(0)}
\label{path}
\end{equation}
where $B(0)=k^4(b-\varepsilon)^2na^2/\varepsilon^2$. Substituting
this expression into Eq.(\ref{path}) and taking into account that
$k=\omega\sqrt{\varepsilon}/c$, we have
\begin{equation}
l\sim
\frac{\varepsilon}{\frac{\omega^2}{c^2}(b-\varepsilon)^2na^2}
\label{path1}
\end{equation}
 Remind that $\varepsilon$ is the
average dielectric constant of the system which for a layered
stack has the form:
\begin{equation}
\varepsilon(\omega)=nab(\omega)+(1-na)\varepsilon_0(\omega)
\label{avcon}
\end{equation}

Here $\varepsilon_0$ is the dielectric constant of a homogeneous
medium into which plates with dielectric constant $b(\omega)$ and
thickness $a$ are randomly embedded. If a homogeneous medium is
vacuum then $\varepsilon_0\equiv 1$. Choosing for plates materials
with $b(\omega)<0$ one can make the average dielectric constant of
the system quite small $\varepsilon\ll 1$. Correspondingly, the
photon elastic mean free path will be small and the radiation
intensity will be large in a such system. Let us now estimate 
the inelastic mean free path $l_{in}$. If the spacing between 
the plates is vacuum then $l_{in}$ is determined by the 
absorption by plates. Absorption in a plate is characterized
by the imaginary part of dielectric constant of a plate
$Imb(\omega)$. Two regimes are possible. When 
$\varepsilon \gg fImb$, where $f$ is the linear fraction of plates
in the system, $l_{in}$ can be estimated as
\begin{equation}
l_{in}\sim \frac{\lambda\sqrt{\varepsilon}}{\pi fImb}
\label{re1}
\end{equation}
In the opposite regime $\varepsilon \ll fImb$, $l_{in}$ is 
estimated as

 \begin{equation}
l_{in}\sim \frac{\lambda}{\pi\sqrt{2fImb}}
\label{re2}
\end{equation}
Substituting Eqs.(\ref{re1}),(\ref{re2}) and Eq.(\ref{path1})
into Eq.(\ref{diff}) one can be convinced that 
$I_D\sim\varepsilon^{-3}$ or $I_D\sim\varepsilon^{-2}$
depending which regime is realized. In any case obviously 
radiation intensity enhances in the limit $\varepsilon\to 0$.
Let us now make some numerical estimates.

In  alkali halide crystals, in semiconductors like $GaP, InSb,$
and etc. the dielectric constant is negative in the region between
the frequencies of transversal and longitudinal optical phonons,
see  \cite{Agr}. For example, for the compound $MgO$ in the
frequency region $550-650cm^{-1}$ the real part of dielectric
constant take values in the interval $-6, -2$ and the imaginary
part in the interval $ 0.6-0.2$. The above mentioned interval lies
in the far infrared region. It follows from Eq.(\ref{path1}) that
in case $2\pi a\leq \lambda$ the minimum of mean free path and
therefore the maximum of radiation intensity is achieved for
average plate thickness $a\sim \lambda/2\pi$. For the above
mentioned frequencies this is about $20\mu m$. Choosing such
values for average thickness of plates one can reach  the
localization limit $\lambda/2\pi$ \cite{And,Arya} which
is a minimal value for photon elastic mean free path $l$.
 One needs $l_{in}$ in order to estimate the number of
emitted photons using Eq.(\ref{diff}). Taking $f\sim
0.1$, $Im b\sim 0.4$ and $\varepsilon\sim 0.5 $, one gets
from Eq.(\ref{re1}), $l_{in}\sim 557\mu m$. Using Eq.(\ref{diff})
one can estimate the integrated over all angles number of emitted
photons in the interval $\Delta \omega$ as
\begin{equation}
N_{ph}\sim \frac{20\pi}{3} \alpha
\left(\frac{l_{in}}{l}\right)^2\frac{\Delta\omega}{\omega}
\label{photon}
\end{equation}
where $\alpha$ is the fine structure constant. Because $l\ll
l_{in}$ the exponential decaying factor in Eq.(\ref{diff}) plays
important role only for very large angles $\theta\approx \pi/2$.
Therefore we ignored it when estimating the total number of
emitted photons. Substituting $l_{in}\sim 562\mu m$, $l\sim
\lambda/2\pi\sim 17\mu m$ into Eq.(\ref{photon})and taking
$\Delta\omega\sim \omega$ one has approximately $N_{ph}\sim 167$
infrared photons per one electron. This implies that using
commercially available $5-6MeV$, $1mA$ linear accelerator a total
output power of $2.4mW( 10^{18}photon/s)$ can be produced.

We have considered radiation effects in systems with near 
zero dielectric constant by example of transition and 
diffusive radiations. An enhancement of intensity is 
demonstrated in both cases. Numerical estimates for specific
systems show that ENZ systems can be used for generation of 
light in the THz and infrared regions.

\acknowledgments
We are grateful to A.Akopian,H.Nersisyan and T.Zaqarian for support.

\end{document}